\documentclass[%
aps,%
prb,%
twocolumn,
showpacs,%
preprintnumbers,%
byrevtex,%
floatfix%
]{revtex4}

\usepackage{epsf}
\usepackage{ifthen}
\usepackage[usenames]{color}
\usepackage{amssymb}
\usepackage{amsfonts}
\usepackage{amsmath}
\usepackage[dvips]{graphicx}
\def\DNA{{\sc dna}}
\def\DNA{{DNA}}
\def\ii{\textrm{i}\,}
\def\ee{\textrm{e}\,}
\def\etal{\textit{et al.}}
\def\ie{\textit{i.e.}}
\def\Hc{\textrm{H.c.}}
\def\eg{\textit{e.g.}}
\def\oc{\omega_{\textrm{c}}}
\def\KB{k_{\textrm{B}}}

\newcommand{\labs} {\left\vert}
\newcommand{\rabs} {\right\vert}
\newcommand{\lsb} {\left[}
\newcommand{\rsb} {\right]}
\newcommand{\lrb} {\left(}
\newcommand{\rrb} {\right)}
\newcommand{\lcb} {\left\{}
\newcommand{\rcb} {\right\}}
\newcommand{\lab} {\left\langle}
\newcommand{\rab} {\right\rangle}

\begin{document}
\date{March,04 2005}

\preprint{\texttt{cond-mat/04?????}}
\title{%
Dissipative Effects in the Electronic Transport through DNA Molecular Wires}
\author{R. Guti{\'e}rrez}
\email{rafael.gutierrez@physik.uni-r.de}
\author{S. Mandal}
\author{G. Cuniberti}

\affiliation{%
Institute for Theoretical Physics, University of Regensburg,
D-93040 Regensburg, Germany}

\begin{abstract}
We investigate the influence of a dissipative environment which effectively
comprises the effects of counterions and hydration shells, on the transport
properties of short \DNA\ wires. Their electronic structure is captured by a
tight-binding model which is embedded in a bath consisting of a collection of harmonic oscillators.
Without coupling to the bath a temperature independent gap opens in the
electronic spectrum. Upon allowing for electron-bath interaction the gap
becomes temperature dependent. It increases with temperature in the weak-coupling limit to the bath degrees
of freedom. In the strong-coupling regime a
bath-induced {\it pseudo-gap} is formed. As a result, a crossover from
tunneling to activated behavior in the low-voltage region of the $I$-$V$
characteristics is observed with
increasing temperature. The temperature dependence of the transmission
near the Fermi energy, $t(E_{\rm F})$, manifests an Arrhenius-like behavior in
agreement with recent transport experiments.
Moreover,
$t(E_{\rm F})$ shows a weak exponential dependence 
on the wire length, typical
of strong incoherent transport. 
Disorder effects smear the electronic bands, but do not appreciably affect the pseudo-gap formation.
\end{abstract}

\pacs{%
87.14.Gg,
87.15.-v 
73.63.-b,
71.38.-k 
72.20.Ee,
72.80.Le, 
05.60.Gg 
}

\maketitle

\section{Introduction}
The idea that conduction pathways in \DNA\ molecules may be built up as a result
of the hybridization of the $\pi$ orbital stack along consecutive base pairs
can be traced back to the 1960's.\cite{spivey62} It was not, however, till
recently that a revival of interest on \DNA\ as a potential conductor occurred.
This was mainly triggered by the observation of long-range electron transfer
between intercalated donor and acceptor centers in \DNA\ molecules in
solution.\cite{murphy93} Subsequent experimental results
\cite{treadway02,meggers98,meggers98a,lewis99,brun94,kelley99,grozema00} were
controversial as they showed different functional dependences of electron
transfer rates on the donor-acceptor separation. Thus, strong exponential
fall-off\cite{meggers98,meggers98a} typical for superexchange mediated transfer
as well as a weak, algebraic dependence \cite{treadway02,kelley99} characteristic of sequential hopping
processes were reported. Meanwhile, theoretical work
has led to an emerging picture where different mechanisms may coexist
depending on base-pair sequence and energetics.\cite{schuster04,jortner98}

In parallel to these developments in the chemical physics community, DC
transport experiments on \mbox{$\lambda$-\DNA} as well as on poly(dG)-poly(dC)
and poly(dA)-poly(dT) molecules between metal electrodes have been
performed.\cite{yoo01,porath00,braun98,kasumov01,storm01,tran00,porath04,endres04} Several
fundamental difficulties have to be surmounted in this kind of experiments: (i)
how to create good contacts to the metal electrodes, (ii) how to control
charge injection into the molecule, (iii) single molecule
vs.~bundles of molecules and (iv) dry vs.~aqueous environments, among others.
Consequently,  sample preparation and the specific experimental conditions turn out to be  very critical 
for DNA transport measurements. Thus, experiments have yielded contradictory results as to
the conduction properties of \DNA\ and are rather difficult to analyze. \DNA\
has been characterized as a pure insulator, \cite{braun98,storm01} as a
wide-band gap semiconductor,\cite{porath00} and as a metallic
system.\cite{yoo01,tao04} Especially interesting are recent transport measurements on 
single poly(dG)-poly(dC) oligomers in aqueous solution, which displayed metallic-like $I$-$V$ 
characteristics and an algebraic 
behavior in the length dependence of the conductance.\cite{tao04} 

Notwithstanding this variety of results and the problems related to the
experimental set-up, the possibility of using \DNA\ in molecular electronics is
extremely attractive since it would open a vast range of potential applications
because of its self-assembling and recognition properties.\cite{braun98}
Alternatively, \DNA\ can also be used as a template in molecular electronic devices.\cite{keren03,hazani04,pompe99}

From a theoretical point of view, the knowledge of the electronic structure of the
base-pairs, the sugar/phosphate mantle and their mutual interactions is
required in order to clarify the transport processes that may be effective in
\DNA. First principle approaches are the most suitable tools for this goal.
However, the huge complexity of this molecule makes {\it ab initio}
calculations still very demanding, so that only few investigations have been
performed, mainly in well-stacked periodic
structures.
\cite{difelice02,felice02,barnett01,gervasio02,artacho03,pablo00,soler03,lewis97,davies04,star04,sankey04}
To complicate this picture, environmental effects such as the presence of water
molecules and counterions which stabilize the molecular structure make {\it ab
initio} calculations even more challenging. \cite{barnett01,gervasio02} Hence,
Hamiltonian
models\cite{hennig04,bishop02,hermon98,yi03,li01,gio02,ao96,zhang04,haenggi05,gmc04}
that isolate single factors affecting
electron transport are still playing a significant role and can help to shed
more light onto the above issues as well as guide first principle investigations.

Recently, Cuniberti \etal~\cite{gio02} proposed a minimal model Hamiltonian to
explain the semiconducting behavior previously observed by Porath
\etal~\cite{porath00} in suspended short (up to 30 base-pairs)
poly(dG)-poly(dC) molecules. Remarkably enough, this experiment was performed
on single molecules, in contrast to most transport experiments involving
bundles of molecules. Molecular systems like Poly(dG)-poly(dC) (or Poly(dA)-poly(dT)) are very attractive from a
theoretical standpoint since, being periodic, band-like transport as a result
of $\pi$-orbitals hybridization may be more efficient than in its strongly
disordered counterparts, \eg~in \mbox{$\lambda$-\DNA}. The above
model\cite{gio02} mimics the electronic structure of the complex
poly(dG)-poly(dC)-backbone system by a tight-binding chain to which side
chains are attached. Electrons can hop along the central chain but not along
the side chains. As a result a gap in the electronic spectrum opens. The gap is
obviously temperature independent and the transmission near the Fermi level
would show a strong exponential dependence due to the absence of electronic
states to support transport.

An immediate issue that arises is how stable this electronic structure, \ie~two
electronic bands separated by a gap, is against the influence of several
factors which are known to play an important role in controlling charge
propagation in \DNA\ molecules, such as static and dynamic
disorder\cite{berlin00,gozema02,roche03,yamada04,ulloa04,unge03,guo04} and
environment.\cite{barnett01,gervasio02,li01} 
 In particular, the environment can act as a
source of decoherence for a propagating electron (or hole),\cite{li01} it can
induce structural fluctuations that support or restrict charge
motion,\cite{barnett01} or it can introduce additional electronic states within
the fundamental gap.\cite{gervasio02,endres04} As it has been
demonstrated experimentally, a modification of the humidity causes variations
of orders of magnitude in the conductivity of \DNA.\cite{jo03,heim04} Moreover, the 
recent single-molecule experiments of Xu \etal~\cite{tao04}  suggest that the environment may 
strongly modified the low-bias transport properties of DNA oligomers.

In this paper we elaborate  on the role played by the environment  by 
addressing 
 signatures of the bath in the electronic transmission spectrum of
the DNA wire in  different coupling regimes: the mean-field approximation as well
as 
weak-coupling and strong-coupling limits. Anticipating some of our
results, we find that the semiconducting gap {\it closes} on the mean-field level as 
a result of thermal fluctuations. In the weak-coupling limit, however, the gap {\it opens}
 with 
increasing temperature. In both cases the electronic gap is an ``intrinsic" property of the
 system. On the contrary, a bath-induced {\it pseudo-gap} is formed in the strong 
coupling 
limit, \ie~an energy region with a low (but finite) density of electronic states. 
 We have further found in this regime  that the transmission at the Fermi
level exponentially decreases with the wire length $L$, $t (E_{\textrm{F}}) \sim
\ee^{-\gamma L}$. The decay rate 
$\gamma$ is however rather small $\sim 0.2$
\textrm{\AA}$^{-1}$. This together with a noticeable dependence of $\gamma$ on the electron-bath 
coupling
clearly indicates that incoherent pathways do appreciably 
contribute to charge transport in the strong coupling limit.

\begin{figure}[tb]
\centerline{\includegraphics[width=.99\linewidth]{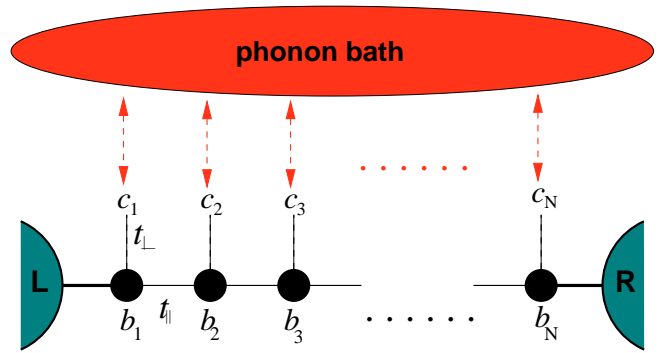}}
\caption{\label{fig:fig1}%
(Color line)  Schematic drawing of the DNA molecular wire in contact with a dissipative environment.
  The central chain with $N$ sites is connected to semiinfinite left (L)
and right (R) electronic reservoirs. 
The bath only interacts  with the side chain sites ({\sf c}), which 
we call  backbone  sites.\cite{comment1}
The Hamiltonian associated with this model is given by
Eqs.~(\ref{eq:eq1}), (\ref{eq:eq2}), and (\ref{eq:eq3}) in the main text.
}
\end{figure}

In the next section we introduce the model Hamiltonian and derive the
corresponding Green functions which are required to calculate the linear
conductance. In section III different approximation schemes associated with
different coupling regimes to the bath are discussed. The influence of
structural disorder on our results is also presented. Finally, our summary
follows in section IV.

\section{Hamiltonian Model}
Along the lines of Refs.~\onlinecite{gio02}, we represent the \DNA\ molecular 
wire containing $N$ base pairs by the following nearest-neighbour 
tight-binding Hamiltonian
(see Fig.~\ref{fig:fig1}):
\begin{eqnarray}
{\cal H}_\textrm{el} &=& \epsilon_b\sum_{j} b^{\dagger}_{j}b_{j} -
t_{||} \sum_{j} \lsb b^{\dagger}_{j}b_{j+1} + \Hc \rsb \nonumber \\ &+& \epsilon \sum_{j}
c^{\dagger}_{j}c_{j} \nonumber \\ &-& t_{\perp}\sum_{j} \lsb
b^{\dagger}_{j}c_{j} + \Hc \rsb \nonumber \\ &=& {\cal H}_\textrm{C} + {\cal
H}_{\textrm{b}} + {\cal H}_{\textrm{C-c}}. \label{eq:eq1}
\end{eqnarray}
Hereby ${\cal H}_{\textrm{C}}$ and ${\cal H}_\textrm{b}$ are the Hamiltonians
of the central and side chains, respectively, and ${\cal H}_{\textrm{C-b}}$ is
the coupling between them. $t_{||}$ and $t_{\perp}$ are hopping integrals along
the central chain and between the backbone sites and the central chain,
respectively. If not stated otherwise, the on-site energies will be later set
equal to zero. The ${\cal H}_{\textrm{C}}$ Hamiltonian can be considered as effectively 
modeling one
of the frontier orbitals of the poly(dG)-poly(dC) system, \eg~the
highest-occupied molecular orbital,  which is localized on the guanine
bases.\cite{artacho03,gervasio02} The side chain induces then a perturbation of the 
$\pi$-stack leading to the opening of a temperature independent 
semiconducting gap in the electronic spectrum, the gap being proportional to 
the transversal hopping integral $t_{\perp}$.~\cite{gio02} Since this model 
shows electron-hole symmetry, two electronic manifolds containing $N$ 
states each, are symmetrically situated around the Fermi level, which is taken as the 
zero of energy.  

We focus here on the influence of the environment on the electronic structure
and consequently on the transport properties of the model described by ${\cal H}_\textrm{el}$. As it has
been demonstrated in the past years, correlated fluctuations of hydrated
counterions strongly influence electron(hole) motion along the
double-helix.\cite{barnett01,endres04} Recent Raman and neutron scattering
experiments on lysozyme have shown that the protein dynamics follows the
solvent dynamics over a broad temperature range. Especially, conformational
changes, low-energy vibrational excitations and the corresponding temperature
dependences turned out to be very sensitive to the solvents dynamics.
\cite{caliskan04} We consider the vibrational degrees of freedom of counterions and hydration shells in
\DNA\ as a dynamical bath able to act as a dissipative environment. In this model Hamiltonian
approach, we do not consider specific features of the environment but
represent it by a phonon bath of $M$ harmonic oscillators.
 We further make the
 assumption that the bath is only directly affecting the side chain
whereas the central chain  is well screened by the latter. Then,
the extended Hamiltonian becomes:
\begin{eqnarray}
{\cal H}_\textrm{W}&=&{\cal H}_\textrm{el} +\sum_{\alpha} \Omega_{\alpha} B^{\dagger}_{\alpha}
B_{\alpha} +\sum_{\alpha,j} \lambda_{\alpha} c^{\dagger}_{j}c_{j}
(B_{\alpha}+B^{\dagger}_{\alpha}) \nonumber \\ &=& {\cal H}_\textrm{el} +{\cal
H}_{\textrm{B}}+{\cal H}_{\textrm{c-B}}, \label{eq:eq2}
\end{eqnarray}
where ${\cal H}_{\textrm{B}}$ and ${\cal H}_{\textrm{c-B}}$ are the phonon
bath Hamiltonian and the backbone-bath interaction, respectively. $B_{\alpha}$ is a  bath phonon 
operator and $\lambda_{\alpha}$ denotes the electron-phonon coupling. Note that we
assume a local coupling of the bath modes to the electronic density at the side
chain. Later on, the thermodynamic limit ($M\to \infty)$ in the bath degrees of freedom will be
carried out and the corresponding bath spectral density introduced, so that at
this stage we do not need to further specify the set of bath frequencies
{$\Omega_{\alpha}$} and coupling constants
{$\lambda_{\alpha}$}.

Finally, we include the coupling of the molecular wire to semiinfinite left (L) and right (R) 
electrodes:
\begin{eqnarray}
{\cal H}&=&{\cal H}_\textrm{W}+\sum_{{\bf k}\in \textrm{L,R}, \sigma} \epsilon_{{\bf
k}\sigma} d^{\dagger}_{{\bf k}\sigma}d_{{\bf k}\sigma} \nonumber \\ &&
+\sum_{{\bf k}\in \textrm{L}, \sigma} ( V_{{\bf k},1} \, d^{\dagger}_{{\bf
k}\sigma} \, b_{1} + \Hc) \nonumber \\ &&+\sum_{{\bf k}\in R, \sigma} (
V_{{\bf k},N} \, d^{\dagger}_{{\bf k}\sigma} \, b_{N} + \Hc) \nonumber \\ &=&
{\cal H}_\textrm{W} + {\cal H}_{\textrm{L/R}}+{\cal H}_{\textrm{L-C}}+{\cal
H}_{\textrm{R-C}} \label{eq:eq3}
\end{eqnarray}

The Hamiltonian of Eq.~(\ref{eq:eq3}) is the starting point of our
investigation. Performing the Lang-Firsov\cite{mahan} unitary transformation $
\bar{\cal{H}} = \ee^{S} {\cal{H}} \ee^{-S} $ with the generator
$S=\sum_{\alpha,j} ( {\lambda_{\alpha}} / {\Omega_{\alpha}}) c^{\dagger}_{j}
c_{j}(B_{\alpha}-B^{\dagger}_{\alpha})$ and $S^{\dagger}=-S$, the linear
coupling to the bath can be eliminated. In the resulting effective Hamiltonian
only the backbone part is modified since the central chain operators $b_\ell $
as well as the leads' operators $d_{\bf{k}\sigma}$ are invariant with respect to the
above transformation. The new Hamiltonian reads:

\begin{eqnarray}
\bar{{\cal H}}&=&{\cal H}_\textrm{C} + {\cal H}_{\textrm{L/R}} + {\cal
H}_\textrm{B} + {\cal H}_{\textrm{L/R-C}} \nonumber \\
&+&(\epsilon-\Delta)\sum_{j} c^{\dagger}_{j}c_{j} - { t_{\perp}\sum_{j} \lsb
b^{\dagger}_{j}c_{j}{\cal X} + \Hc \rsb } \label{eq:eq4} \\ {\cal X}&=&
\exp{\lsb \sum_{\alpha}
\frac{\lambda_{\alpha}}{\Omega_{\alpha}}(B_{\alpha}-B^{\dagger}_{\alpha})\rsb},
\; \; \; \Delta = \sum_{\alpha} \frac{\lambda_{\alpha}^2}{\Omega_{\alpha}}.
\nonumber
\end{eqnarray}

Let's define two kinds of retarded thermal Green functions related to the
central chain $G_{j \ell}(t)$ and to the backbones $P_{j \ell}(t)$,
respectively ($\hbar=1$):
\begin{eqnarray}
\label{eq:eq5}
G_{j \ell}(t)&=&-i\Theta(t)\lab \lsb b_{j}(t),b^{\dagger}_{\ell}(0)\rsb _{+}
\rab, \\ P_{j \ell}(t)&=&-i \Theta(t) \lab \lsb c_{j}(t){\cal
X}(t),c^{\dagger}_{\ell}(0){\cal X}^{\dagger}(0)\rsb _{+} \rab, \nonumber
\end{eqnarray}
where $\Theta$ is the Heaviside function and the average is taken w.r.t. $\bar{{\cal H}}$. With the above definitions and using
the equation of motion technique (see Appendix \ref{appendix:app1}) 
we arrive to an expression for
the Fourier transform of the central chain Green function which reads, to 
lowest-order in $t_{\perp}$:
\begin{eqnarray}
{\bf G}^{-1}(E)&=&{\bf G}_0^{-1}(E)
- t^2_{\perp} {\bf P}(E) \label{eq:eq6} \\ {\bf G}_0^{-1}(E)&=& E{\bf 1}-{\cal
H}_\textrm{C}-\Sigma_{\textrm{L}}(E)- \Sigma_{\textrm{R}}(E). \nonumber
\end{eqnarray}
In this equation ${\bf G}_0 (E)$ is the Green function of a chain without
backbones and connected to the left and right electrodes. The influence of
the latter is comprised in the complex self-energy functions
$\Sigma_{\textrm{L/R}}(E)$.\cite{datta_book}
The polaronic Green function ${\bf P}(E)$ is explicitly given by:

\begin{eqnarray}
P_{\ell j}(E)&=& -\ii \delta_{\ell j}\int_{0}^{\infty} \mathrm{d}t \,
\ee^{{\ii}(E+{\ii}0^{+})t}\, \ee^{-{\ii}(\epsilon-\Delta)\,t} \, \nonumber \\
&& \times \left [(1-f_{\sf c}) \ee^{-\Phi(t)} + f_{\sf c} \ee^{-\Phi(-t)} \right ]
\label{eq:eq7}
\end{eqnarray}

with $\ee^{-\Phi(t)}= \lab{\cal X}(t) {\cal X}^{\dagger}(0)\rab_{\textrm{B}}$
being a dynamical bath correlation function. The average $\lab \cdot
\rab_{\textrm{B}}$ is performed over the bath degrees of freedom. Working to 
lowest order in $t_{\perp}$ allows to use a zero-order Green function for the side chain 
in Eq.~(7), \ie~
$G^{\sf c}_{0,\ell j}(t)\sim  \delta_{\ell j} \ee^{-{\ii}(\epsilon-\Delta)\,t}$.
$f_{\sf c}$ is the Fermi 
function at the
backbone sites. In what follows we consider the case of  empty sites by setting $f_{\sf c}=0$. Note that
${\bf P}$ is a diagonal matrix, \ie~it only modifies the on-site energies in
the Hamiltonian.

In order to get closed expressions for the bath thermal averages it is
appropriate to introduce a bath spectral density\cite{weiss_book} defined by :
\begin{eqnarray}
J(\omega)= \sum_{\alpha} \lambda^{2}_{\alpha} \delta(\omega-\Omega_{\alpha}) = J_0
( \frac{\omega}{\oc})^s \ee^{-\omega/\oc} \Theta(\omega),
\label{eq:eq8}
\end{eqnarray}
where $\oc$ is a cut-off frequency related to the bath memory time
$\tau_\textrm{c}\sim \oc^{-1}$. It is easy to show that the limit $\oc\to
\infty$ corresponds to a Markovian bath, \ie~ $J(t)\sim J_0 \delta(t)$. Using this {\it Ansatz}, $\Phi(t)$ can
be written in the usual way:\cite{weiss_book}
\begin{eqnarray}
\Phi(t)&=& \int_{0}^{\infty} \mathrm{d}\omega \, \frac{J(\omega)}{\omega^2}\lsb
1-\ee^{-{\ii}\omega t}+2\frac{1-\cos{\omega t}}{\ee^{\beta\omega}-1}\rsb .
\label{eq:eq9}
\end{eqnarray}
Although the integral can be performed analytically\cite{weiss_book}, we 
consider $\Phi(t)$ in some limiting cases where it is easier to work
directly with Eq.~(\ref{eq:eq9}).

In the transport calculations, we limit ourselves to treat the low voltage regime,
thus neglecting  non-equilibrium
effects as well as the inelastic part of the total current. 
As a result, one can still define a linear conductance 
$g$  as follows:~\cite{imry04} 
\begin{eqnarray}
g(E)&=&\frac{2e^2}{h}\int dE \, \lrb  -\frac{\partial f}{\partial E} \rrb \, t(E),  \\
\label{eq:eq10}
t(E)&=&{\textrm{Tr}} \lcb
{\bf \Delta_{{\rm L}}}\,
\mathbf{G}\,   
{\bf \Delta_{{\rm R}}} \,
\mathbf{G}^\dagger
\rcb, \nonumber
\end{eqnarray}
where 
${\bf \Delta_{\rm L,R}}={\rm i}\left ( \mathbf{\Sigma}_{\rm L,R}- \mathbf{\Sigma}^{\dagger}_{\rm L,R}\right ) $
 are 
spectral densities of the leads.
Although the foregoing expression is similar to Landauer's formula, we stress
that the influence of the phonon bath does implicitly appear via the
Green function ${\bf G}$. Hence, both coherent {\it and} incoherent pathways for
charge transport mediated by phonon processes are included in Eq.~(10). We concentrate our
discussion on the temperature and length dependence of $t(E)$. In what follows we always plot $t(E)$ rather
than $g$ to filter out temperature effects arising from the derivative of the Fermi function in
Eq.~(10). For completeness
the current as given by 
$I(V)=( {2e }/ {h}) \int \mathrm{d}E \, (f(E-{eV}/{2})-f(E+{eV}/{2})) t(E)$ 
is also shown.
We remark however, that this expression neglects non-equilibrium effects, which
are beyond the scope of this investigation.

\section{Limiting cases}
We use now the results of the foregoing section to discuss the electronic
transport properties of our model in some
limiting cases for which analytic expressions can be derived. In all cases, we
use the wide-band limit in the electrode selfenergies,
\ie~$\Sigma_{\textrm{L},\ell j}(E)=-{\ii}\Gamma_{\textrm{L}}\delta_{1
\ell}\delta_{1j}$ and $\Sigma_{\textrm{R},\ell
j}(E)=-{\ii}\Gamma_{\textrm{R}}\delta_{N \ell}\delta_{Nj}$. 
We discuss the mean-field approximation and  the weak-coupling regime in
the electron-bath interaction as well as  the strong-coupling limit. 
 Farther, the cases of ohmic ($s=1$) and  superohmic ($s=3$) spectral densities are
 treated. 
\begin{figure}[t]
\centerline{\includegraphics[width=.99\linewidth]{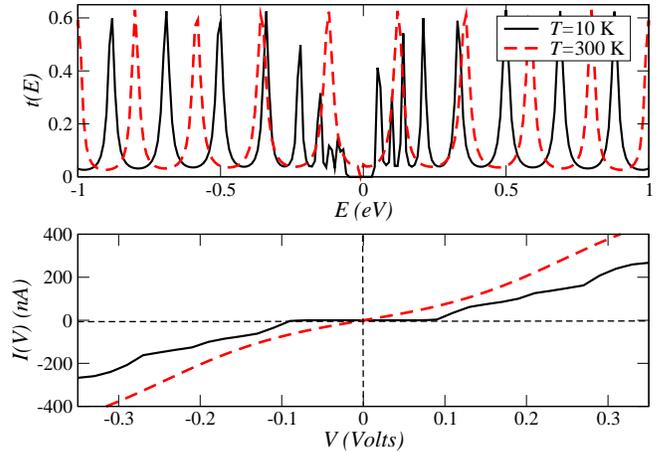}}
\caption{\label{fig:fig2}%
(Color online) Electronic transmission and corresponding current in the mean-field
approximation for two different temperatures. Parameters: $N=20, J_0/\oc= 0.12,
t_{\perp}/t_{||}=0.5, \Gamma_{\textrm{L/R}}/t_{||}=0.5$.
}
\end{figure}

\subsection{Mean-field approximation (MFA) }
Within the mean-field approximation bath fluctuations contained in $P(E)$ are neglected. 
The MFA can be introduced by writing the phonon operator
${\cal X}$ as $\lab{\cal X}\rab_{\rm B} +\delta {\cal X}$ in ${\cal
H}_{\textrm{C-c}}$ in Eq.~(\ref{eq:eq4}), \ie~${\cal
H}^{\textrm{MF}}_{\textrm{C-c}}=-t_{\perp}\sum_{j} \lsb
b^{\dagger}_{j}c_{j}\lab{\cal X}\rab_{\textrm{B}} + \Hc \rsb + O(\delta {\cal
X})$. As a result a real, static and temperature dependent term
in Eq.~(\ref{eq:eq6}) is found:
\begin{eqnarray}
{\bf G}^{-1}(E)={\bf G}_0^{-1}(E)- t^2_{\perp} \frac{|\lab {\cal
X}\rab_{\textrm{B}}|^{2}}{E-\epsilon+\Delta+\ii 0^{+}}{\bf 1},
\label{eq:eq11}
\end{eqnarray}
where $\labs \lab {\cal X}\rab_\textrm{B} \rabs^2=\ee^{-2\kappa(T)}$ and
$\kappa(T)$ is given by:
\begin{eqnarray}
\kappa(T)= \int_{0}^{\infty} \frac{\mathrm{d}\omega}{\omega^2} J(\omega)
\coth{\frac{\omega}{2 \KB T}}. 
\end{eqnarray}
The effect of the MF term is thus to scale the bare transversal hopping
$t_{\perp}$ by the exponential temperature dependent factor $\ee^{-\kappa(T)}$.

In the case of an ohmic bath, $s=1$, the integrand in $\kappa(T)$ scales as $1/\omega^{p},\,
p=1,2$ and 
has thus a logarithmic divergence
at the lower integration limit, see Eqs.~(\ref{eq:eq8}) and (12). Thus,
the MF contribution would vanish. In other words, no gap would
exist on this approximation level. 

In the superohmic case ($s=3$) all integrals are regular. One obtains 
$\Delta=\int \mathrm{d}\omega \,
\omega^{-1} J(\omega)= \Gamma(s-1)J_0=2J_0$, with $\Gamma(s)$ being the Gamma
function and $\kappa(T)$ reads:
\begin{eqnarray}
\kappa(T) = & \nonumber \\ & \frac{2J_0}{\oc} \left [ \, 2 \lrb
\frac{\KB T}{\oc}\rrb ^{2} \, \zeta_{\textrm{H}} \lrb 2,\frac{\KB
T}{\oc}\rrb -1 \right ].  \label{eq:eq13}
\end{eqnarray}
$ \zeta_{\textrm{H}}(s,z)=\sum_{n=0}^{\infty} (n+z)^{-s}$ is the Hurwitz
$\zeta$-function, a generalization of the Riemann
$\zeta$-function.\cite{gradshteyn_book}

It follows from Eq.~(13) that $\kappa(T)$ behaves like a constant
for low temperatures ($\KB T/\oc < 1$), $\kappa(T)\sim J_0/\oc$, while it
scales linear with $T$ in the high-temperature limit ($\KB T/\oc > 1$),
$\kappa(T)\sim J_0/\oc(1+2 \KB T/\oc)$.

If $J_0$ vanishes, $\Delta$ is zero and $\lab{\cal X}\rab_{\textrm{B}}=1$. Thus
we recover the original model of Ref.~\onlinecite{gio02} which has a gap
proportional to $t_{\perp}$. For $J_0\neq 0$ and at zero temperature the
hopping integral is roughly reduced to $t_{\perp}\ee^{-\frac{J_0}{\oc}}$ which
is similar to  the renormalization of the hopping in Holstein's
polaron model,\cite{holstein59} though here it is $t_{\perp}$ rather than $t_{||}$ the 
 term that is 
rescaled. At high temperatures $t_{\perp}$ is further
reduced ($\kappa(T)\sim T$) so that the gap in the electronic spectrum 
finally collapses and the system becomes metallic, see Fig.~\ref{fig:fig2}. An
appreciable temperature dependence can only be observed in the limit $J_0/\oc <
1$; otherwise the gap would collapse already at zero temperature due to the
exponential dependence on $J_0$. We further remark that the MFA 
may be only valid in the
regime $J_0/\oc < 1$, $\KB T/\oc \lesssim 1$, otherwise 
 multiphonon processes in the bath, which are not
considered at this stage, become increasingly relevant.

\subsection{Beyond MF: weak-coupling limit}
As a first step beyond the mean-field approach let's 
first consider the weak-coupling limit in ${\bf P}(E)$. For $J_0/\oc <
1$ and not too high temperatures ($\KB T/\oc \lesssim 1$) the main contribution
to the integral in Eq.~(\ref{eq:eq7}) comes from long times $t\gg \oc^{-1}$.
\begin{figure}[t]
\centerline{\includegraphics[width=.99\linewidth]{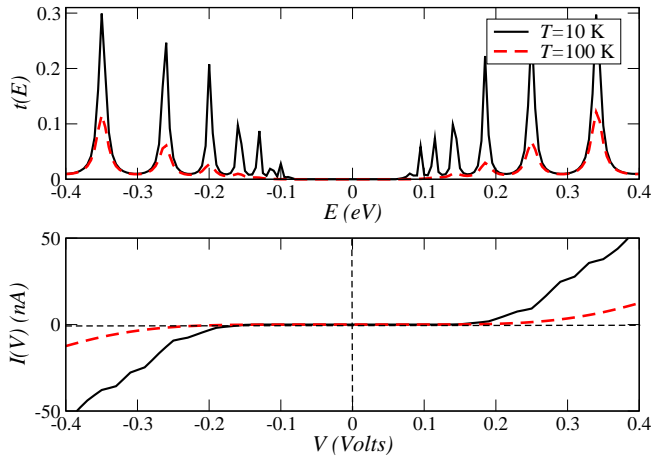}}
\caption{\label{fig:fig3}%
(Color line) Electronic transmission and corresponding current in the weak-coupling limit
with ohmic dissipation ($s=1$) in the bath. Parameters: $N=20, J_0/\oc= 0.2,
t_{\perp}/t_{||}=0.6, \Gamma_{\textrm{L/R}}/t_{||}=0.5$
}
\end{figure}
With the change of variables $z=\omega t$, $\Phi(t)$ can be written as:
\begin{eqnarray}
\Phi(t)&=&J_0 \oc^{-s} t^{1-s} \int_{0}^{\infty} \mathrm{d}z \, z^{s-2} \ee^{-\frac{z}{\oc
t}} \nonumber \\ & &\times \left (1-\ee^{-\ii z}+
2\frac{1-\cos{z}}{\ee^{z\frac{\beta\oc}{\oc t}}-1}\right). \label{eq:eq14}
\end{eqnarray}
As far as $\oc t \gg \beta\oc $ this can be simplified to:

\begin{eqnarray}
\Phi(t)&\approx & J_0 \oc^{-s} t^{1-s} \int_{0}^{\infty}  \mathrm{d}x \, z^{s-2}
\ee^{-\frac{z}{\oc t}} \nonumber \\ & & \times \lrb 1-\ee^{-\ii z}+
2\frac{\beta\oc}{\oc t}\frac{1-\cos{z}}{z} \rrb. \label{eq:eq14}
\end{eqnarray}
Since in the long-time limit the low-frequency bath modes are giving the most
important contribution we may expect some qualitative differences in the ohmic
and superohmic regimes. For $s=1$ we obtain $\Phi(t)\sim \pi
\frac{J_0}{\oc}\frac{\KB T}{\oc}(\oc t)$ which leads to (using $\Delta(s=1)=J_0$):
\begin{eqnarray}
{\bf G}^{-1}(E)&=&{\bf G}_0^{-1}(E)- t^2_{\perp} \frac{1}{E+J_0+\ii
\pi\frac{J_0}{\oc} \KB T}{\bf 1}, \label{eq:eq15}
\end{eqnarray}
\ie~there is only a pure imaginary contribution from the bath. For the simple case of
a single site coupled to a backbone one can easily see that the gap
approximately scales as $\sqrt{\KB T}$; thus it grows with increasing
temperature. This is shown in Fig.~\ref{fig:fig3}, where we also see that the
intensity of the transmission resonances strongly goes down with increasing
temperature. The gap enhancement is induced by the suppression of 
the transmission peaks of the frontier orbitals, i.~e.~those closest to the Fermi energy. 

For $s=3$ and $\KB T/\oc \lesssim 1$, $\Phi(t)$ takes a nearly temperature
independent value proportional to $J_0/\oc$. As a result the gap is
slightly reduced ($t_{\perp}\to t_{\perp}\ee^{-J_0/\oc}$) but, because of the
weak-coupling condition, the effect is rather small. \\ From this discussion
we can conclude that in the weak-coupling limit ohmic dissipation in the bath
induces an enhancement of the electronic gap while superohmic dissipation does not
appreciably affect it. In the high-temperature limit $\KB T/\oc > 1$ a
short-time expansion can be performed which yields similar results to those
of the strong-coupling limit (see next section),\cite{ao96} so that we do
not need to discuss them here. Note farther that the gap obtained 
in the weak-coupling and mean-field limits 
is an ``intrinsic'' property of the electronic system; it is only quantitatively modified by the interaction 
with the bath degrees of freedom. We thus trivially 
expect a strong exponential dependence of $t(E=E_{{\rm F}})$ on the wire length, typical
of virtual tunneling through a gap. Indeed, we find $t(E=E_{{\rm F}})\sim \exp{(-\beta \, L)}$ with 
$\beta\sim 2-3 \, \AA^{-1}$.

\begin{figure}[t]
\centerline{
\includegraphics[width=.90\linewidth]{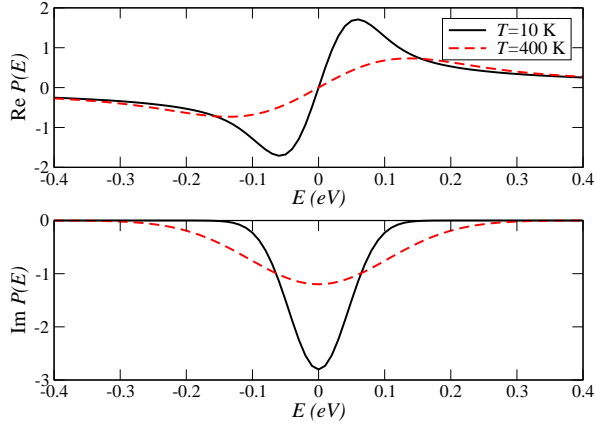}%
}
\caption{\label{fig:fig4}
(Color line) Temperature dependence of the real and
imaginary parts of $P(E)$ for $N=20$, $J_0/\oc=10, t_{\perp}/t_{||}=0.4,
\Gamma_{\textrm{L/R}}/t_{||}=0.5$. With increasing temperature the slope of the
real part near $E=0$ decreases and the imaginary part broadens and loses
intensity. A similar qualitative dependence on $J_0$ was found (not shown).
}
\end{figure}

\subsection{Beyond MF: strong coupling limit (SCL)}
In this section we discuss  the strong-coupling regime, as defined by the 
condition $J_0/\oc > 1$. This may be the regime to be found in presence of  
an aqueous environment, as recent theoretical estimations using the classical Onsager 
model for solvation processes have shown.~\cite{gilmore05}  In
the SCL the main contribution to the time integral in Eq.~(\ref{eq:eq7}) arises
from short times. Hence a short-time expansion of $\Phi(t)$ may already give
reasonable results and it allows, additionally, to find an analytical
expression for ${\bf P}(E)$. At $t\ll \oc^{-1}$ we find,
\begin{eqnarray}
\label{eq:eq16} \Phi(t)&\approx&{\ii} \Delta \, t + (\oc t)^2 \, \kappa_{0}(T) \\
P_{\ell j}(E)&=&-{\ii}\delta_{\ell j}\, \, \int_{0}^{\infty} \mathrm{d}t \,
\ee^{{\ii}(E-\epsilon+{\ii}0^{+})t}\, \ee^{-(\oc t)^2 \kappa_{0}(T)} \nonumber \\
&=&-{\ii}\delta_{\ell j}\, \, \frac{\sqrt{\pi}}{2}\frac{1}{\oc
\sqrt{\kappa_{0}(T)}} \; \exp \lrb
{-\frac{(E-\epsilon+{\ii}0^{+})^2}{4\omega^2_\textrm{c} \kappa_{0}(T)}} \rrb
\nonumber \\ && \times \lrb 1+{\textrm{erf}}\lsb
\frac{{\ii}(E-\epsilon+{\ii}0^{+})}{2\oc \sqrt{\kappa_{0}(T)}}\rsb \rrb, \nonumber \\
\kappa_{0}(T)&=& \frac{1}{2\omega^2_\textrm{c}}\int_{0}^{\infty} \mathrm{d}\omega J(\omega)
\coth{\frac{\omega}{2 \KB T}}.  \nonumber
\end{eqnarray}

Before presenting the results for the electronic transmission, it is
useful to first consider the dependence of the real and imaginary parts of
${\bf P}(E)$ on temperature and on the reduced coupling constant $J_0/\oc$.
Both functions are shown in Fig.~\ref{fig:fig4}. We see that around the Fermi level at 
$E=0$ the
real part is approximately linear, ${\rm Re} \, P(E)\sim E$ while the imaginary part shows a
Gaussian-like behavior. The imaginary part loses intensity and becomes
broadened with increasing temperature or $J_0$, while the slope in the real
part decreases when $\KB T$ or $J_0$ are increased.
\begin{figure}[t]
\centerline{
\epsfclipon
\includegraphics[width=.99\linewidth]{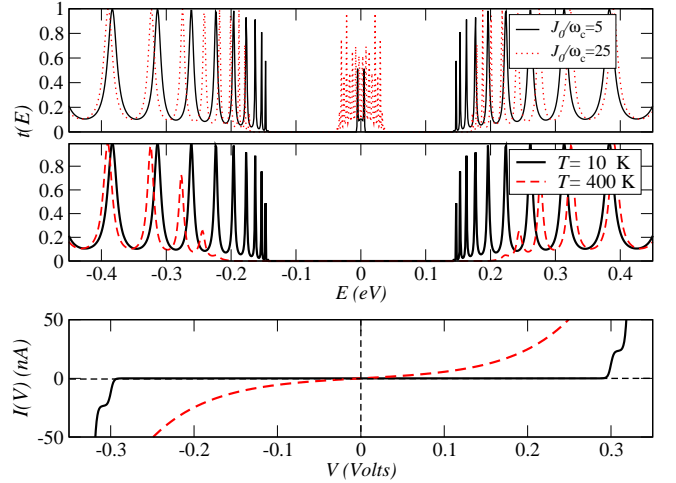}%
\epsfclipoff
}
\caption{\label{fig:fig5}%
(Color line)  Upper panel: $t(E)$ with ${\rm{Im}}\, P(E)=0$; the
intensity of the resonances on the central narrow band is strongly dependent on
$J_0/\omega_c$ and $\KB T$ (not shown). Temperature dependence of $t(E)$ with full 
inclusion of $P(E)$
(middle panel) and corresponding current (lower panel) for $N=20$, $J_0/\oc=5,
t_{\perp}/t_{||}=0.5, \Gamma_{\textrm{\textrm{L/R}}}/t_{||}=0.2$. The
pseudo-gap increases with temperature.
}
\end{figure}
If we neglect for the moment the imaginary part (the dissipative influence of
the bath), we can understand the consequences of the real part being nonzero
around the Fermi energy, \ie~in the gap region.
 The solutions of the non-linear equation
${\rm{det}}|(E-t_{\perp}^{2} {\rm Re}\, P(E)){\bf 1}-{\cal H}_\textrm{C}|=0$ give the new
poles of the Green function of the system  in presence of the phonon bath.
For comparison, the equation determining the eigenstates {\it without} the bath
is simply ${\rm{det}}|(E-t_{\perp}^{2}/E){\bf 1}-{\cal H}_\textrm{C}|=0$. It
is just the $1/E$ dependence near $E=0$ that induces the appearance of two
electronic bands of states separated by a gap.\cite{gio02} In our present
study, however, ${\rm Re} \, P(E\to 0)$ has no singular behavior and additional
poles of the Green function  may be expected to appear in the low-energy sector. 
This is indeed the
case, as shown in Fig.~\ref{fig:fig5}\ (upper panel). We find a third band of
states around the Fermi energy, which we call a polaronic band because it results from the 
strong interaction between an electron and the bath modes. The  intensity of this band as well as its band 
width strongly depend
on temperature and on $J_0$. When $\KB T$ (or $J_0$) become large enough, these
states spread out and eventually merge with the two other side bands. This
would result in a transmission spectrum similar of a metallic system. 

This picture is nevertheless not complete since the imaginary component of
$P(E)$ has been neglected. Its inclusion leads to a dramatic modification of
the spectrum, as shown in Fig.~\ref{fig:fig5}\ (middle panel). We now only see
two bands separated by a gap which basically resembles the semiconducting-type
behavior of the original model. The origin of this gap or rather
{\it pseudo-gap} (see below) is however quite different. It turns out that the
imaginary part of $P(E)$, being peaked around $E=0$, 
strongly suppresses the transmission resonances
belonging to the central band. Additionally, the frontier orbitals on the side bands, \ie~ orbitals closest 
to the gap region, are also strongly damped, this
effect becoming stronger with increasing temperature (${\rm{Im}}\,P(E)$
broadens). This latter effect has some similarities with the previously discussed 
weak-coupling regime. 
Note, however, that the new electronic manifold around the Fermi energy 
does not appear in the weak-coupling
regime. We further stress that the density
of states around the Fermi level is not exactly zero (hence the term
pseudo-gap); the states on the polaronic  manifold, although strongly damped,
contribute nevertheless with a finite, temperature dependent incoherent background to the
transmission. As a result, with increasing temperature, a crossover from
tunneling to activated behavior in the low-voltage region of the $I$-$V$
characteristics takes place, see Fig.~\ref{fig:fig5}\ (lower panel). The slope
in the $I$-$V$ plot becomes larger when $t_{\perp}$ is reduced, since the side
bands approach each other and the effect of ${\rm{Im}}\,P(E)$ is reinforced. 

\begin{figure}[t]
\centerline{\includegraphics[width=.99\linewidth]{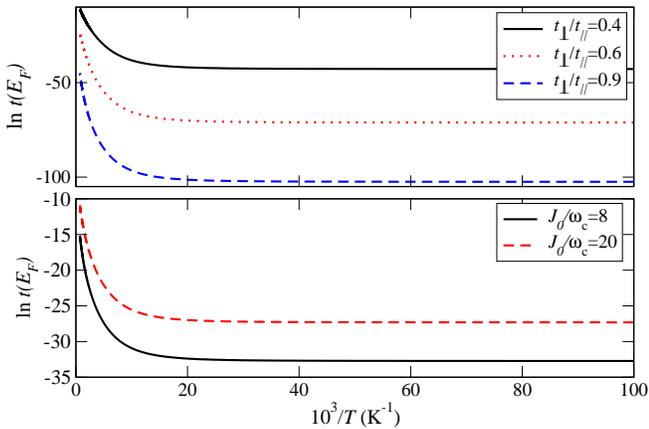}}
\caption{\label{fig:fig6}%
(Color line)  Arrhenius plot of  $t(E=E_{{\rm F}})$ for different
 transversal couplings $t_{\perp}$ (upper panel) and electron-bath couplings $J_0/\oc$
 (lower
 panel). Parameters: $N=20, t_{\perp}/t_{||}=0.5,
 \Gamma_{\textrm{\textrm{L/R}}}/t_{||}=0.25$.
 }
 \end{figure}

In Fig.~\ref{fig:fig6} an Arrhenius plot of the transmission at the Fermi energy
is shown for different strengths of the transversal hopping integral and the 
electron-bath coupling. After a nearly $T$-independent region, the transmission strongly grows
up following approximately a $\ee^{-1/T}$ law. Increasing the coupling to
the phonon bath makes the suppression of the polaronic  band around $E=0$ less
effective (${\rm{Im}}\,P(E\sim 0)$ decreases) so that the density of
states around this energy becomes larger. Hence the absolute value of the
transmission also increases. Similar $T$-dependences have been
experimentally observed in poly(dG)-poly(dC)\cite{yoo01} as well as in
\mbox{$\lambda$-\DNA}.\cite{tran00} On the other side, increasing $t_{\perp}$ 
leads to a reduction of the 
transmission at the Fermi level, since the energetic separation of the side
 bands increases with  $t_{\perp}$.

\begin{figure}[b]
\centerline{
\includegraphics[width=.99\linewidth]{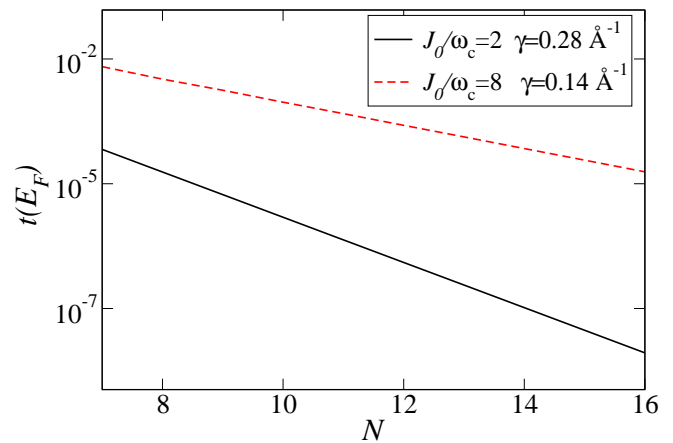}
}
\caption{\label{fig:fig7}%
(Color line)  Chain length dependence of the transmission function at the Fermi
energy for different electron-bath interaction strengths. Parameters: 
$t_{\perp}/t_{||}=0.125,
\Gamma_{\textrm{\textrm{L/R}}}/t_{||}=0.15, T=200 \,{\rm K} $.
}
\end{figure}

We have further investigated the length
dependence of the transmission at the Fermi energy. This is a very important aspect that 
helps to identify the influence of different transport mechanisms.\cite{jortner98,segal00}
The results are displayed
in Fig.~\ref{fig:fig7}\ for different values of the reduced coupling $J_0/\oc$.
For a homogeneous chain (on-site energies are set to zero) an exponential
dependence on the chain length $t(E_{\textrm{F}})\sim \ee^{-\gamma L}$ was
found. In this expression $L=N a_0$, where $N$ is the number of sites on the molecular wire and
$a_0 \sim 3.4 \, {\rm \AA }$ is the average distance between consecutive base pairs. 
Note that the inverse decay lengths $\gamma$ are rather small $\sim 0.1-0.3 \, \rm{\AA}^{-1}$.
An exponential dependence usually indicates virtual tunneling through
a gap. Inverse decay lengths, as extracted e.~g.~from complex band structure
calculations,~\cite{sankey04,fagas04} are however much larger that those 
obtained in the present
investigation. 
So have recent DFT-based calculations found values of $\gamma\sim 1.5 \, \rm{\AA}^{-1}$ 
for gap tunneling in dry Poly(dG)-Poly(dC) oligomers.\cite{sankey04}
 With increasing bath coupling the
exponential dependence farther weakens and eventually becomes algebraic
$t(E_{\textrm{F}})\sim N^{-\alpha}$. 
The
introduction of a tunnel barrier as realized \eg~through insertion of (AT)$_n$
groups, by shifting  the on-site energies along a finite segment of the chain
increases the inverse decay length $\gamma$  by a factor of 2, approximately.
Obviously, this model cannot describe the crossover from superexchange mediated
electron transfer (strong exponential behavior) to sequential hopping-mediated
transport (algebraic dependence) as a function of the {\em wire length} $N$, as discussed in other
works.\cite{jortner98,segal00} We guess that vibrational excitations inside the
central chain, which renormalize the longitudinal hopping integral $t_{||}$, 
 have to be included to get this non-monotonic transition.

 From the previous discussion we may conclude that electron transport on the low-energy sector of the 
transmission spectrum is supported by the formation of  polaronic states. Though strongly damped, these 
states manifest  nonetheless with  
a finite density of states inside the bandgap.

It has been meanwhile demonstrated \cite{roche03,yamada04,barnett01,gozema02,ulloa04,unge03,guo04} that
electron (or hole) motion in \DNA\ is extremely sensitive to different kinds of disorder: static
disorder (random base-pair sequences), structural fluctuations and inhomogeneities of the
counterions distribution along the backbones. These factors may strongly distort
the base pair stacking along
the double helix and eventually affect the electronic transport properties. They deserve
a separate study. However,
as a test for the stability of our results we
have randomly varied the on-site energies along the central chain
by extracting them from a Gaussian distribution with variance $\sigma_0$. In this way we are
simulating some kind of structural disorder induced, \eg~by thermal fluctuations inside the
central chain. In Fig.~8 the cases of weak ($\sigma_0\sim 0.12 t_{||}$) and strong disorder ($\sigma_0\sim
t_{||}$) are shown. Two main features can be seen: (i) the transmission resonances on the
side bands are strongly washed out and lose in
intensity, and (ii) the pseudo-gap is slightly reduced with increasing disorder.
However, the suppression
of the central band due to ${\rm{Im}} \, P(E)$ and hence, the pseudo-gap formation is not affected by
this kind of disorder. As soon as electronic states shift from the side bands into the region with
nonzero ${\rm{Im}} \, P(E)$ they
are strongly damped and thus the pseudo-gap structure of the spectrum is conserved. 
A similar effect of disorder is expected in the other coupling regimes to the bath degrees 
of freedom discussed above.

\section{Summary}
Charge propagation in \DNA\ molecules is extremely sensitive to disorder and
environmental effects. We have focused in this paper on the influence of a
dissipative environment on the electronic transport properties of a 
model Hamiltonian which mimics some basic features of the electronic structure
of \DNA\ oligomers.  Although we have chosen
Poly(dG)-Poly(dC) molecules as a reference point, we believe that our model is quite 
generic and may be useful for  a large class of $\pi$-conjugated systems.

\begin{figure}[t]
\centerline{\includegraphics[width=.99\linewidth]{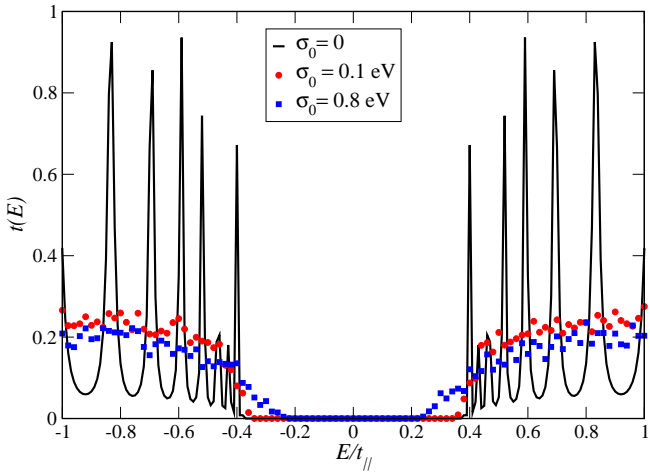}}
\caption{(Color line)
 Transmission function in presence of thermal disorder in the central chain.
Parameters: $N=20, J_0/\oc=5, t_{\perp}/t_{||}=0.5,
\Gamma_{\textrm{\textrm{L/R}}}/t_{||}=0.15, T=10 {\rm K} $. The transmission on the
side bands decreases when the disorder becomes stronger, but the pseudo-gap is
still seen, although it is partially reduced with increasing disorder.
}
\end{figure}

We have shown that a mean-field
approximation cannot fully catch the action of a dissipative environment on
charge transport, because it only gives a real, energy independent
contribution. Indeed, while the mean-field approach leads to gap reduction
with increasing temperature, bath fluctuations eventually
lead to gap opening in the weak-coupling limit. 
We have further shown that a
bath-induced pseudo-gap in the electronic spectrum can appear for strong
electron-bath coupling giving a temperature-dependent background around the
Fermi energy. As a result the system may show with increasing temperature a
transition from a tunneling to an activated behavior in the low-bias region
when coupled to an external dissipative bath. An Arrhenius-like temperature
dependence of the transmission at the Fermi level and a rather weak exponential
dependence on the wire length were additionally found, indicating a strong contribution 
of incoherent pathways of the charge carriers.

A natural extension of this investigation would be the inclusion of
non-equilibrium effects at large bias and consequently of inelastic components
of the current. This issue is although interesting from a formal point of view,
since the Lang-Firsov transformation introduces polaronic
rather than pure electronic propagators, see Eq.~(5). For the former the appropriate Keldysh
Green functions should be derived in order to deal with the non-equilibrium
regime. This problem deserves a separate investigation which is now in
progress.\\

\section{Acknowledgments}
We would like to thank M. Hartung and J. Keller  for fruitful
discussions. This work has been supported by the Volkswagen foundation and by
the EU under contract IST-2001-38951. \\

\begin{appendix}
\section{\label{appendix:app1}Derivation of Eq.~(\ref{eq:eq6})}
The equation of motion for the retarded Green function in Eq.~(\ref{eq:eq5}) in
the frequency representation reads:
\begin{eqnarray}
E G_{\ell j}(E)=\lab \lsb b_j,b_\ell \rsb _{+} \rab +(( \lsb b_j,\cal{\bar{H}}\rsb |b_\ell
)).\nonumber
\end{eqnarray}
Using it we get for the Hamiltonian of Eq.~(\ref{eq:eq3}) :
\begin{eqnarray}
&&\sum_{n}\lsb G^{-1}_{0}(E)\rsb _{\ell n} G_{nj}(E)= \delta_{\ell
j}-t_{\perp}((c_\ell {\cal X}|b^{\dagger}_{j})) 
\label{eq:eqA1}
\end{eqnarray}
\vspace{-0.5cm}
\begin{eqnarray}
\Big[G^{-1}_{0}(E)\Big]_{\ell n} &=& (E-\epsilon_b)\delta_{n
\ell}+t_{||}(\delta_{n, \ell+1}+\delta_{n, \ell-1}) \nonumber \\ & &-
\Sigma_{\textrm{L}}\delta_{\ell 1}\delta_{n1}-\Sigma_{\textrm{R}}\delta_{\ell
N}\delta_{nN} \nonumber \\ \Sigma_{\textrm{L(R)}}&=&\sum_{{\bf k}\in L(R)}
\frac{|V_{{\bf k},1(N)}|^2}{E-\epsilon_{{\bf k}}+\ii 0^{+}} \nonumber
\end{eqnarray}
Now,
equations of motion from the ``right'' may be written for the Green function
$Z_{\ell j}^{{\cal X}}(E)=((c_\ell {\cal X}|b^{\dagger}_{j}))$,

 leading to :
\begin{eqnarray}
\sum_{m} Z_{\ell m}^{{\cal X}}(E) \lsb G^{-1}_{0}(E)\rsb _{mj}&=&-t_{\perp}
((c_\ell {\cal X}|c^{\dagger}_{j}{\cal X}^{\dagger}))\nonumber \\
&=&-t_{\perp} P_{\ell j}(E) \label{eq:eqA2} 
\end{eqnarray}
In the former equations we have neglected cross-terms of the form $((c_\ell
{\cal X}|c^{\dagger}_j))$, since they will give contribution of $O(t^3_{\perp})$. Inserting Eq.~(\ref{eq:eqA2}) 
into Eq.~(\ref{eq:eqA1}) we arrive
at the matrix equation:
\begin{eqnarray}
{\bf G}(E)={\bf G}_{0}(E)+ {\bf G}_{0}(E) {\bf \Sigma}_{\textrm{B}}(E) {\bf
G}_{0}(E), \nonumber
\end{eqnarray}
which can be transformed into a Dyson-like equation when introducing the
irreducible part ${\bf \Sigma}_{\textrm{B}}(E)={\bf
\Sigma}^{\textrm{irr}}_{\textrm{B}}(E)+{\bf
\Sigma}^{\textrm{irr}}_{\textrm{B}}(E){\bf G}_{0}(E) {\bf
\Sigma}^{\textrm{irr}}_{\textrm{B}}(E)+\dots$:
\begin{eqnarray}
{\bf G}(E)={\bf G}_{0}(E)+ {\bf G}_{0}(E) {\bf
\Sigma}^{\textrm{irr}}_{\textrm{B}}(E) {\bf G}(E). \label{eq:eqA4}
\end{eqnarray}
From Eq.~(\ref{eq:eqA4}) it immediately follows Eq.~(\ref{eq:eq6}) with ${\bf
\Sigma}^{\textrm{irr}}_{\textrm{B}}(E)=t^2_{\perp} {\bf P}(E)$.
We emphasize that these expressions are exact only to lowest-order in the 
 transversal hopping $t_{\perp}$. This approximation may be justified in the low-voltage 
 limit we are dealing with.

\end{appendix}


\begin{thebibliography}{63}
\bibitem[{com()}]{comment1}
\bibinfo{note}{We speak of backbones in a very generic way. In our model the
  side chain (=backbones) is representative of the sugar/phosphate mantle as
  well as of the complementary strand. Hence, electronic states on the
  backbones (in our use of the term) are not necessarily localized on the
  sugar/phosphate molecular groups.}
\end{thebibliography}
\end{document}